**Human services organizations and the responsible integration of AI: Considering ethics and contextualizing risk(s)**

Brian E. Perron[1], Lauri Goldkind[2], Zia Qi[1], and Bryan G. Victor[3]

[1] School of Social Work, University of Michigan – Ann Arbor

[2] Graduate School of Social Service, Fordham University

[3] School of Social Work, Wayne State University

**Author Note**

Correspondence concerning this article should be directed to Brian E. Perron, University of Michigan - Ann Arbor, 1080 South University Avenue, Ann Arbor, MI 48109, United States. E-mail: beperron@umich.edu

Abstract

This paper examines the responsible integration of artificial intelligence (AI) in human services organizations (HSOs), proposing a nuanced framework for evaluating AI applications across multiple dimensions of risk. The authors argue that ethical concerns about AI deployment—including professional judgment displacement, environmental impact, model bias, and data laborer exploitation—vary significantly based on implementation context and specific use cases. They challenge the binary view of AI adoption, demonstrating how different applications present varying levels of risk that can often be effectively managed through careful implementation strategies. The paper highlights promising solutions, such as local large language models, that can facilitate responsible AI integration while addressing common ethical concerns. The authors propose a dimensional risk assessment approach that considers factors like data sensitivity, professional oversight requirements, and potential impact on client wellbeing. They conclude by outlining a path forward that emphasizes empirical evaluation, starting with lower-risk applications and building evidence-based understanding through careful experimentation. This approach enables organizations to maintain high ethical standards while thoughtfully exploring how AI might enhance their capacity to serve clients and communities effectively.

Human services organizations and the responsible integration of AI: Considering ethics and contextualizing risk(s)

The adoption of artificial intelligence (AI) tools in human services organizations has been hindered by a lack of resources and a scarcity of evidence-informed practices (Goldkind, et al, 2024). Human services organization (HSO) leaders may also be skeptical of implementing AI tools that have been criticized as environmentally hazardous, inherently biased and developed using exploitative and extractive practices. The reticence to engage in the AI discourse and a lack of timely and accurate information obscures the complex reality that AI applications exist along a spectrum of risk ranging from benign to catastrophic, with consequential ethical implications. The lack of nuance impedes thoughtful evaluation of the potential benefits and drawbacks of implementing these new tools in social service delivery.

Through analysis of implementation contexts, evidence from real-world deployments, and an examination of ethical implications, this paper provides suggestions for HSOs to evaluate AI applications across multiple dimensions of risk. We contend that ethical concerns manifest differently depending on data sensitivity, potential impact on client wellbeing, professional oversight requirements, and the types of AI tools being used. Highlighted here are emerging solutions such as open-source local large language models that can help to facilitate responsible AI integration.

To start, we explore four commonly cited ethical concerns around the deployment of AI within HSOs: outsourcing professional judgment and decision-making, energy consumption and environmental depletion, model bias and the exploitation of data laborers in AI development; and how these ethical concerns vary in significance based on contextualizing factors (Li et al., 2022; Maphosa, 2024). This dimensional analysis reveals how the severity and manageability of each ethical concern shift based on specific AI implementation choices, such as model size, implementation approach, and integration with professional practice. Understanding these variations enables organizations to make context-specific decisions about AI adoption, by situating the AI deployment in context.

This dimensional approach to analyzing AI risks aligns with recent U.S. Department of Health and Human Services (2024) guidelines, which similarly advocates for a tiered framework that distinguishes between AI applications that are presumed to impact rights and safety, those that may impact rights and safety depending on implementation, and those unlikely to impact rights and safety. Similar to those guidelines, we emphasize that the severity of ethical concerns shifts based on specific implementation choices and contexts, enabling organizations to calibrate oversight and safeguard appropriately to the level of risk. Organizations should also recognize that some applications of AI in the human services should be categorically avoided, particularly those involving automated decision-making about critical services, care plans, or interventions without meaningful human oversight. For example, using AI to independently determine child welfare placements, automate benefits eligibility decisions, or generate clinical diagnoses would create unacceptable risks to client wellbeing and violate core social work ethics.

Moving beyond polarized positions on AI adoption, we hope to foster more constructive discussions about how HSOs can thoughtfully engage with AI technologies while maintaining core principles of equity, access, and social justice. We conclude by proposing a path that emphasizes empirical evaluation, starting with lower-risk applications and building evidence-based understanding through careful experimentation with local language models. This measured approach allows the field to maintain its high ethical standards while thoughtfully exploring how AI might enhance its capacity to serve clients and communities effectively.

**Ethical Issues in Context**

The path to responsible AI integration in human services requires unpacking common ethical concerns that often lead to blanket rejection of AI technologies. The following section examines four key ethical issues frequently arising in AI adoption discussions: the potential displacement of professional judgment, energy consumption and environmental impact, model bias, and harms to data labelers. We argue that risk manifests differently across various applications for each concern, challenging the notion that all AI implementations carry equal ethical weight.

**Outsourcing of Professional Judgment and Decision-Making**

The ethical concern about AI replacing professional judgment represents the most fundamental challenge to AI adoption in social work. Critics rightly worry that automated systems might diminish professional autonomy or substitute algorithmic decisions for human judgment in client care (Creswell-Baez et al., under review). These concerns have been particularly acute in child welfare, with jurisdictions increasingly making predictive AI analytics available to workers in decision-making (Trail, 2024). We share these concerns and hold that humans -- not AI-powered tools -- should always remain the active, ethical agent in practice decisions. Our analysis suggests that the risk of outsourcing professional judgment to AI models exists across applications, with some technologies potentially enhancing rather than replacing professional judgment.

For example, retrieval augmented generation (RAG) systems demonstrate how carefully constrained AI can augment professional capabilities while preserving autonomy (Perron et al., in press; Victor et al., 2024). These systems combine a knowledge base containing verified information with a language model that processes user queries. When a professional queries the system about organizational policies or procedures, it retrieves relevant sections from official documentation and uses these specific passages to construct its response. Every answer can be traced back to its source material, making verification straightforward and maintaining accountability. This structured approach ensures professionals retain decision-making authority while gaining more efficient access to institutional knowledge.

The key to addressing this ethical concern lies in how AI tools are positioned within professional practice. Rather than automated decision-making, these systems serve as information assistants that help professionals work more efficiently. When reviewing case documentation, AI can highlight relevant passages and suggest potential connections to organizational resources, but the professional maintains complete control over interpretation and decision-making. Similarly, while AI can help categorize and summarize information, professionals continue to exercise their judgment in determining how to use that information in client care.

This evidence-based understanding transforms the ethical concern from fear of replacement to thoughtful consideration of augmentation. By carefully selecting applications that support rather than replace professional judgment, organizations can maintain high standards of practice while improving efficiency in information access and routine tasks. Like our analysis of other ethical concerns, this reveals how thoughtful implementation can strengthen rather than compromise professional autonomy.

**Energy Consumption and Environmental Impact**

ChatGPT, Claude, and Gemini, among others, are f*rontier cloud-based AI* models representing the commercial forefront of generative artificial intelligence. These models are characterized by their cutting-edge architecture, scalability, and ability to handle large-scale, complex tasks (Anderljung, et al, 2023). They are massive in scale, measured in "parameters," and they are model-specific coefficients or instructions that give the model direction, whose exact numbers remain closely guarded trade secrets. The computational demands and resource requirements of these tools are substantial, particularly regarding energy consumption (Li, et al, 2024).

The environmental impact of these models (Ren, et al, 2024) has become a critical concern for our profession, especially given our commitment to environmental justice (Minnick et al., 2024). The training process alone requires enormous computational resources, translating into significant energy consumption. This impact is well-documented. For example, BLOOM, a 176-billion parameter language model, produced approximately 50.5 tonnes of $CO_2$ emissions—equivalent to about 60 transatlantic flights between London and New York (Luccioni et al., 2022).

Fortunately, the landscape of AI technology has expanded beyond resource-intensive frontier models to include more sustainable alternatives. For example, practical applications can use smaller AI models that run on local computers with far less environmental impact. These smaller models - called 'local' or 'open-source' models - are designed to run efficiently on standard computers without requiring cloud connections or specialized hardware. Unlike their larger counterparts, these models are typically free to download and use, and many are optimized for specific tasks like text classification or document analysis. They can operate entirely within an organization's IT infrastructure, which is more energy efficient and substantially increases data security. Small local models can run disconnected from the Internet, addressing privacy concerns and data leaks. While local models may not match the broad capabilities of frontier models, they can perform as effectively for specific use cases using only a fraction of the computational resources (Alassan et al., 2024).

To put this efficiency in perspective, while ChatGPT-4o contains 1.76 trillion parameters, the open-source Llama3-8B model uses only 8 billion parameters while achieving human-level accuracy on text classification tasks (Perron et al., 2024). The waffle chart visualizes this difference, with each square representing 8 billion parameters (one Llama3 model). This 220-fold difference in scale demonstrates the dramatic contrast in architectural approaches between these models.

**[INSERT FIGURE 1 ABOUT HERE]**

Our research group routinely processes and accurately categorizes batches of 50,000 case notes using the open-source Llama-8B, which is run on a secure, local server, in just 48 hours. In contrast, manual review would require about 2,000 hours at an average review speed of 2 minutes per record. Completing a business-as-usual manual review thus uses orders of magnitude more energy via months of continuous computing operations—likely across multiple workstations—while the AI model completes the work in two days on a single workstation.

While concerns about the energy consumption of frontier AI models are significant and real, focusing solely on large-scale systems overlooks the potential environmental and organizational benefits of utilizing smaller, local models for specific tasks. By thoughtfully selecting and implementing the right-sized AI solutions, organizations can reduce their computational footprint and achieve significant environmental and resource gains over traditional manual processes. This balanced approach to AI adoption acknowledges the challenges and opportunities in responsible technology implementation.

**Model Bias**

The challenges of model bias in high-stakes decision-making are well-documented (Gallegos, et al, 2024). Model bias manifests when AI systems perform differently or make systematically different recommendations across demographic groups, potentially perpetuating or amplifying societal inequities. For example, research has shown that large language models sometimes generate harmful, race-based medical advice when provided with patients' demographics (Omiye et al., 2023). Similarly, AI-powered facial recognition systems used in security and law enforcement consistently show higher error rates for Black people (Bacchini & Lorusso, 2019).

The risk of model bias can vary significantly, depending on the task the AI model is performing. Natural language processing (NLP) tasks like classification (categorizing text into predefined groups), extraction (pulling specific information from documents), and summarization (condensing text while maintaining key points) have a fundamentally different risk profile than creative content generation. These structured tasks operate on existing content rather than generating new ideas, making them less susceptible to bias and fabrication.

The key advantage of these NLP tasks is that their accuracy can be evaluated using the same methods we use to assess human performance. Unlike creative or predictive applications, these structured operations have clear, objective criteria for success: The classification aligns with expert judgment, extracted information either matches the source document or doesn't, and the summary either captures the key points or misses them. Even in these lower-risk scenarios, organizations can implement safeguards such as establishing clear quality thresholds before deployment, regularly auditing samples of AI-processed work against human review, maintaining diverse validation datasets representing all client populations, and ensuring transparent documentation of where AI tools are being used.

**Harms to Data Labelers**

Most AI products depend on training data, which are used to "teach" the model how to generate outputs. Training data is often delivered to a software developer "raw" or under-described and requires a cleaning and labeling process to be useful (Huang & Zhao, 2024). Data labeling in AI refers to annotating data with meaningful tags or labels that help machine learning (ML) models learn patterns from the data during training. During the labeling process, annotators are often exposed to disturbing content, including violence, abuse, and graphic imagery, causing psychological trauma and stress disorders (Crawford, 2023). Content moderators and data labelers frequently report anxiety, depression, and PTSD symptoms from reviewing harmful material (Strongylou, et al, 2024). This creates an ethical burden where AI development relies on human psychological harm.

However, recent advances in synthetic data generation offer compelling alternatives that can eliminate the need for human exposure to traumatic content while maintaining model effectiveness and also increasing representation in training data (Kang, et al, 2024). For example, Microsoft's development of the Phi-3 model family demonstrates how synthetic data generation can avoid human labeling (Beatty, 2024). Their approach began with carefully curated educational content, using larger models to generate training data that maintained the clarity and structure of textbook explanations. This process, called "CodeTextbook," eliminated the need for human annotators to review potentially harmful content. Instead, the synthetic data was generated through an iterative process where quality educational materials were used to create new training examples, filtered, and refined through automated processes.

The development of synthetic data generation techniques, as demonstrated by Microsoft's Phi-3 model family and similar initiatives, represents a significant shift in how AI models are being trained without exposing human annotators to harmful content. This evolution in training methodologies illustrates how blanket criticisms of AI ethics often fail to account for rapid technological advances that address these concerns. While vigilance about AI ethics remains important, the field's evolution toward safer, effective alternatives demonstrates the need for context-specific evaluation of AI applications.

## A continuum of low- to high-stakes AI applications

The path to responsible AI integration in human services requires nuanced consideration of how different applications affect client well-being, professional judgment, and organizational effectiveness. While the field has seen growing interest in AI applications across practice areas, the risk profile of each application varies along multiple dimensions that must be carefully evaluated. These dimensions include the immediacy and magnitude of potential impact on client outcomes, the degree of professional oversight possible, and the sensitivity of data being processed.

Applications involving direct client care decisions that require nuanced professional judgment and a deep understanding of complex human situations present the highest risk profiles. Treatment planning systems, risk prediction algorithms that guide intervention decisions, and AI-driven therapy without meaningful professional oversight all operate at the higher end of the risk spectrum. The limitations of current AI technology and the profound consequences of errors in these domains make such applications particularly challenging to implement responsibly. The inability of AI systems to fully grasp the context, account for intersecting social factors, or understand human experience requires extensive safeguards and may not be appropriate for automation with current technology.

Toward the lower end of the continuum are applications with progressively more manageable risk profiles, particularly those focusing on administrative and organizational tasks. These applications demonstrate how risk can be mitigated through system design choices: outputs can be verified before affecting clients, professional judgment remains firmly in control, and operations focus on organizational rather than client data. Rather than viewing these as "low-stakes," organizations should evaluate each application across multiple risk dimensions to determine appropriate safeguards and implementation approaches. For instance, even seemingly routine tasks like document classification require careful consideration of data privacy, potential bias in categorization, and the downstream effects on service delivery.

We propose examining every potential AI application through a dimensional risk assessment framework that considers data sensitivity, professional oversight requirements, and potential impact on client well-being. This nuanced approach enables organizations to make context-specific decisions about implementation and identify appropriate safeguards based on where each application falls along various risk dimensions. Rather than asking whether an application is "high-stakes" or "low-stakes," we encourage organizations to ask, "What kinds of risks does this application present, can we manage them effectively, and if so, what steps will we take to ensure ethical implementation?"

To illustrate how organizations might apply our dimensional risk framework, we present examples of tasks that typically present more manageable risk profiles due to their administrative nature and built-in verification opportunities. While these tasks often fall toward the lower end of the risk continuum, their actual risk level depends heavily on organizational context, implementation approach, and specific use case. These applications can be designed with important risk-mitigation characteristics: utilizing small language models that run locally on organizational hardware to address environmental impact and data privacy concerns, incorporating explicit human oversight through structured verification processes to maintain professional control over outputs, and focusing on augmenting rather than replacing professional judgment to create more time for direct client engagement. However, even these seemingly routine applications require careful evaluation across our risk dimensions - verification timelines, data sensitivity, oversight requirements, and potential client impact - to determine appropriate safeguards for working within each organization's context.

**Document Analysis and Classification**

- Analyzing historical case notes, records, or reports to identify patterns and classify content for research, evaluation, or planning purposes where findings can be verified against source materials and validated through sampling before influencing decisions.

**Information Access and Organization**

- Retrieving relevant sections from organizational policies, procedures, and documentation in response to staff queries where every response can be verified against source materials.
- Organizing and indexing institutional knowledge bases to improve information accessibility while maintaining clear links to authoritative sources.

**Administrative Data Processing**

- Cleaning and standardizing datasets for analysis where outputs can be validated against raw data using established quality control procedures.
- Generating routine visualizations and reports from administrative data where results can be verified through standard statistical validation practices.

**Documentation Support**

- Identifying potential connections between cases and available organizational resources while leaving interpretation and decision-making to professionals.
- Highlighting relevant passages in lengthy documents to support professional review while maintaining human control over interpretation.

**Resource Management**

- Analyzing historical resource utilization patterns to support planning where recommendations can be validated before implementation.
- Identifying potential inefficiencies in administrative workflows where suggested changes can be evaluated before adoption.

**Records Management**

- Organizing and categorizing archived records for improved accessibility where classifications can be verified against original documents.
- Identifying duplicate or inconsistent records where changes can be reviewed before being implemented.

**Training Material Development**

- Processing existing training materials to create indexed resources where all content comes from approved sources.
- Generating practice scenarios from de-identified historical cases where materials can be reviewed before use.
- Synthesize peer-reviewed academic literature on evidence-based practice for use in professional development contexts.

**Quality Assurance Support**

- Checking documentation against standardized requirements where discrepancies can be verified manually.
- Identifying potential documentation gaps where findings can be validated before any corrective action.

**Research Support**

- Analyzing de-identified datasets for research purposes where findings can be validated through established scientific methods.
- Extracting relevant information from published literature where results can be verified against source materials.

**Operational Monitoring**

- Tracking administrative metrics and generating alerts about operational issues where warnings can be verified before action is taken.
- Analyzing workflow patterns to identify potential bottlenecks where findings can be validated before process changes.

    We encourage organizations to use these examples as inspiration rather than rigid prescriptions. Local contexts, resources, and needs will shape how different applications manifest in practice. By sharing implementation experiences and outcomes across the field, we can collectively better understand where AI can responsibly enhance practice in human services. This evidence-based approach to expanding AI applications aligns with the field's commitment to careful evaluation and continuous improvement in service delivery.

**Moving the Field Forward**

Our analysis of ethical concerns and risk dimensions in AI adoption reveals several critical paths forward for human services research and practice. The field needs coordinated effort across three key areas to advance responsible AI integration: implementation research, professional education, and organizational capacity building.

The first priority is developing standardized frameworks for evaluating AI implementations in human services contexts. While our dimensional analysis provides a starting point, we need empirical research on how organizational contexts affect risk profiles and implementation outcomes. This research should focus on understanding how verification processes, oversight mechanisms, and professional integration practices influence implementation success across different positions on the risk continuum. Particular attention should be paid to studying how organizations successfully implement verification processes for administrative applications, as these findings could inform the development of more robust safeguards for higher-risk applications.

Professional education represents the second critical area for advancement. Public administration, nonprofit administration and social work programs need to integrate AI literacy into their curricula, focusing not on technical skills but on critical evaluation capabilities. Students and practitioners need frameworks to assess AI applications across our identified risk dimensions and understand how different implementation choices affect ethical concerns. This education should emphasize hands-on experience with local language models in controlled environments, allowing practitioners to develop a practical understanding of both capabilities and limitations while learning to identify appropriate use cases within their specific practice contexts.

The third priority is building organizational capacity for AI evaluation and implementation, which will likely require considerable investment in staff development and organizational learning. Organizations need structured processes and policy guidance to assess potential AI applications across our identified risk dimensions and access to comprehensive training programs to develop AI literacy among leadership, technical staff, and frontline workers. This includes developing internal expertise in local model deployment, establishing verification protocols, creating feedback mechanisms to track implementation outcomes, and ensuring staff understand AI tools' capabilities and limitations. Clear policy guidance and governance frameworks are also essential, particularly around data privacy, model testing requirements, and criteria for human oversight. Given these various challenges, organizations should start by implementing applications at the lower end of the risk continuum, using their experiences to build institutional knowledge about effective oversight mechanisms and integration practices before considering more complex applications. Professional associations and regulatory bodies will also need to play a role in developing standards and providing guidance similar to that of HHS, to ensure consistent and responsible AI adoption across the sector.

These priorities point toward a concrete research agenda. We need case studies documenting successful implementations of administrative AI applications, focusing particularly on how organizations develop effective verification processes and professional oversight mechanisms. Research should examine how different organizational contexts affect implementation outcomes and what factors contribute to successful integration with professional practice. This evidence base will be crucial for developing more detailed guidelines for responsible AI adoption in social work settings.

By focusing on these priorities, the field can move beyond theoretical debates to evidence-based evaluation of AI applications in human services. This approach allows us to adhere to our values while thoughtfully exploring how AI might enhance our capacity to serve clients and communities more effectively.

HUMAN SERVICES ORGANIZATIONS AND AI                                                                 19

HUMAN SERVICES ORGANIZATIONS AND AI                                                                 20Creswell-Báez, J., Victor, B.G. , Dysart, C., & Goldkind, L. (under review). "I don't understand it, but okay": An empirical study of mental health practitioners' readiness to use large language models. *Journal of Technology in Human Services.*

Gallegos, I. O., Rossi, R. A., Barrow, J., Tanjim, M. M., Kim, S., Dernoncourt, F., ... & Ahmed, N. K. (2024). Bias and fairness in large language models: A survey. *Computational Linguistics*, 1-79. https://doi.org/10.1162/coli_a_00524

Goldkind, L., Ming, J., & Fink, A. (2024). AI in the nonprofit human services: Distinguishing between hype, harm, and hope. *Human Service Organizations: Management, Leadership & Governance*, 1-12. https://doi.org/10.1080/23303131.2024.2427459

Huang, Q., & Zhao, T. (2024). Data collection and labeling techniques for machine learning. *arXiv preprint arXiv:2407.12793 [cs.DB].* https://doi.org/10.48550/arXiv.2407.12793

Kang, A., Chen, J. Y., Lee-Youngzie, Z., & Fu, S. (2024). Synthetic data generation with LLM for improved depression prediction. *arXiv preprint arXiv:2411.17672 [cs.LG].* https://doi.org/10.48550/arXiv.2411.17672

Li, F., Ruijs, N., & Lu, Y. (2023). Ethics & AI: A systematic review on ethical concerns and related strategies for designing with AI in healthcare. *AI, 4*(1), 28-53. https://doi.org/10.3390/ai4010003

Li, Y., Mughees, M., Chen, Y., & Li, Y. R. (2024). The unseen AI disruptions for power grids: LLM-induced transients. *arXiv preprint, arXiv:2409.11416 [cs.AR].* https://doi.org/10.48550/arXiv.2409.11416

Luccioni, A. S., Viguier, S., & Ligozat, A-L. (2022). Estimating the carbon footprint of BLOOM, a 176B parameter language model. *arXiv Preprint, arXiv:2211.02001 [cs.LG].* https://arxiv.org/abs/2211.02001

HUMAN SERVICES ORGANIZATIONS AND AI                                                                 21Maphosa, V. (2024). The rise of artificial intelligence and emerging ethical and social concerns. *AI, Computer Science and Robotics Technology, 3*(1), 1–20. https://doi.org/10.5772/acrt.20240020

Minnick, D., Rao, S., Smith, K., Krings, A., & Teixeira, S. (2024). *Meeting the grand challenge to create social responses to a changing environment.* American Academy of Social Work & Social Welfare. https://grandchallengesforsocialwork.org/wp-content/uploads/2024/07/240730-SRTCE_PB_FINAL.pdf

Molala, T. S., & Zimba, Z. F. (2023). Ethically-sound artificial intelligence for potential deployment in social work education and practice: A global south perspective. *Proceedings of the Fourth Southern African Conference for Artificial Intelligence Research.* https://2024.sacair.org.za/wp-content/uploads/2023/11/9-sacair23.pdf

Omiye, J. A., Lester, J. C., Spichak, S., Rotemberg, V., & Daneshjou, R. (2023). Large language models propagate race-based medicine. *NPJ Digital Medicine, 6*(1), 195. https://doi.org/10.1038/s41746-023-00939-z

Perron, B. E., Luan, H., Victor, B. G., Hiltz-Perron, O., & Ryan, J. (2024). Moving beyond ChatGPT: Local large language models (LLMs) and the secure analysis of confidential unstructured text data in social work research. *Research on Social Work Practice.* Advance online publication. https://doi.org/10.1177/10497315241280686

Perron, B. E., Hiltz, B. S., Khang, E. M., & Savas, S. A. (in press). AI-enhanced social work: Developing and evaluating retrieval augmented generation (RAG) support systems. *Journal of Social Work Education.*

**Figure 1:**

**Comparative Scale Visualization of Large Language Model Parameters**

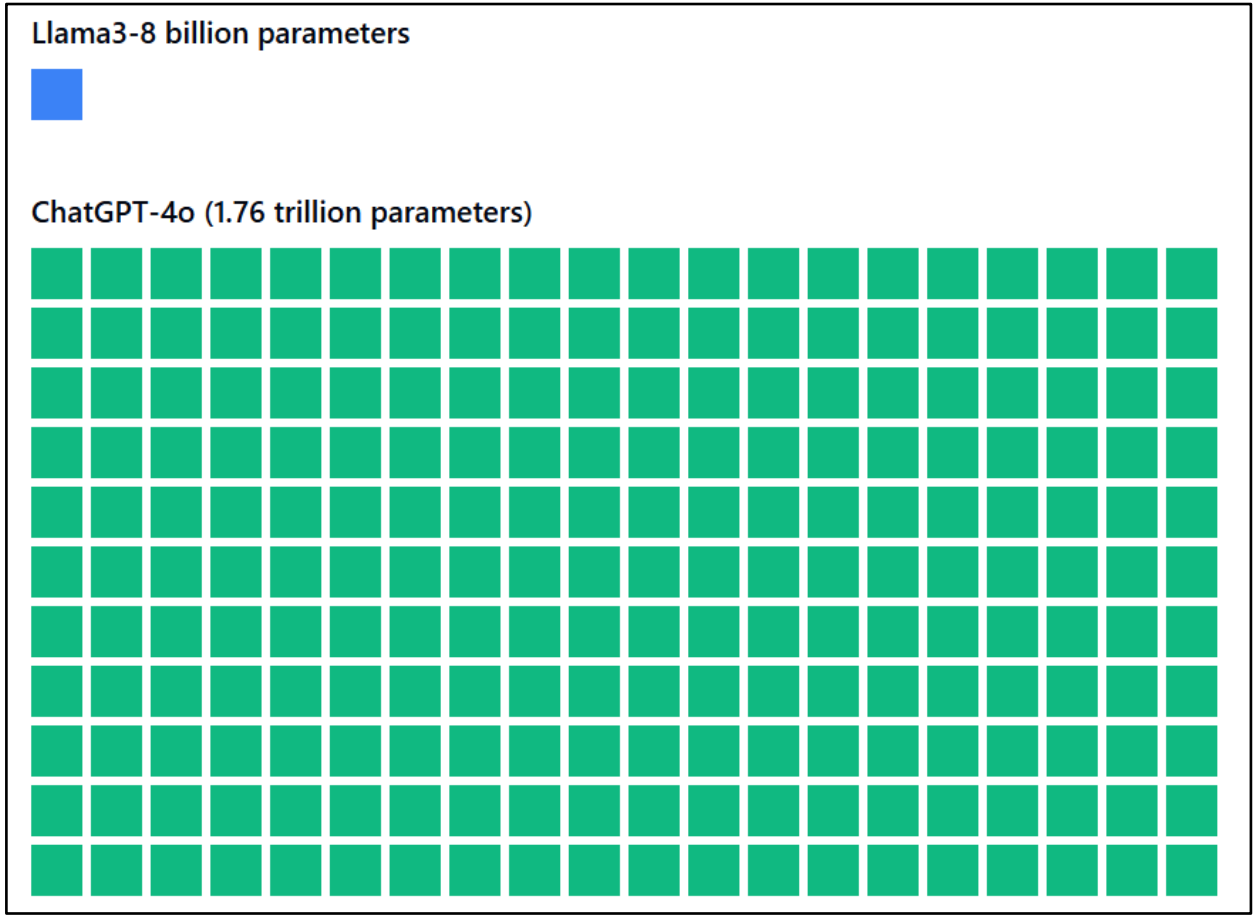

Note: Each square represents 8B parameters (equivalent to one Llama3-8B model). The number of squares for ChatGPT-4o is a conservative estimate.